\documentclass[reprint,prl]{revtex4-1}

\pdfoutput=1

\usepackage[italian,english]{babel}
\usepackage{amssymb, amsmath}
\usepackage{graphicx}
\usepackage{mathrsfs,eufrak}

\newcommand{\beq}{\begin{equation}}
\newcommand{\eeq}{\end{equation}}
\newcommand{\beqm}{\begin{multline}}
\newcommand{\eeqm}{\end{multline}}
\newcommand{\bdis}{\begin{displaymath}}
\newcommand{\edis}{\end{displaymath}}
\newcommand{\bea}{\begin{eqnarray}}
\newcommand{\eea}{\end{eqnarray}}
\newcommand{\barr}{\begin{array}}
\newcommand{\earr}{\end{array}}
\newcommand{\bfig}{\begin{figure}[!]}
\newcommand{\efig}{\end{figure}}

\makeatletter
\let\cat@comma@active\@empty
\makeatother

\begin{document}
\title{Electric field effect on spin waves and magnetization dynamics: \\ role of magnetic moment current}

\author{Vittorio Basso}
\affiliation{Istituto Nazionale di Ricerca Metrologica, Strada delle Cacce 91, 10135 Torino, Italy}
\author{Patrizio Ansalone}
\affiliation{Istituto Nazionale di Ricerca Metrologica, Strada delle Cacce 91, 10135 Torino, Italy}

\date{\today}

\begin{abstract} We show that a static electric field $E_x$ gives rise to a shift of the spin wave dispersion relation $\omega(q_y-q_E)$ in the direction of the wavenumber $q_y$ of the quantity $q_E=-\gamma_LE_x/c^2$. This effect is caused by the magnetic moment current carried by the spin wave itself that generates an additional phase proportional to the electric field, as in the Aharonov-Casher effect. This effect is independent from the possibly present magneto-electric effects of insulating ferromagnets and superimposes to them. By extending this picture to arbitrary magnetization dynamics, we find that the electric field gives rise to a dynamic interaction term which has the same chiral from of the Dzyaloshinskii-Moriya interaction but is fully tunable with the applied electric field. \end{abstract}
 

\maketitle

The understanding of the physical basis of the transport of magnetic moment in the solid state is the central issue of spintronics where the information is expected to be carried by the spin instead of the charge \cite{Zutic-2004}. In insulating ferromagnets the magnetic moment current, or spin current, is due to spin waves, the excitation of the magnetization field \cite{Serga-2010}. Spin waves can be easily generated, transmitted and detected, but some method to manipulate their phase is expected in order to be used in magnonics interference devices \cite{Lenk-2011}. For example spin waves have been shown to acquire a phase when they traverse a non uniform magnetization configuration \cite{Hertel-2004, Dugaev-2005}. In this context the possibility to achieve a fine tuning of the acquired phase by a static electric field has been already the subject of several research efforts. The energy of non centrosymmetric multiferroics, like BiFeO$_3$, having a spontaneous polarization, has been shown to include a magneto-electric coupling term \cite{Mostovoy-2006} that can provide an electric control of spin waves \cite{Rovillain-2010, Risinggaard-2016}. The extension to centrosymmetric crystals, in which the electric polarization is induced by an electric field, was considered by Mills and Dzyaloshinskii \cite{Mills-2008} and by Liu and Vignale \cite{Liu-2011}. As a consequence of the magneto-electric energy term, the spin wave dispersion relation $\omega(q)$ is modified by an additional term which is linear in the wavenumber and is proportional to the electric field, as $\omega(q)+\omega_MbE_xq_y$, where $b$ is the strength of the magneto-electric coupling. The idea stimulated several recent developments \cite{Wang-2018, Krivoruchko-2018} aiming to a detailed development of electric field controlled phase shifters.

However, beside these magneto-electric effects which modifies the energy, the magnetic moment current (or spin current) transported by the spin wave itself modifies the linear momentum. In presence of a static electric field, $\mathbf{E}$, the canonical linear momentum of a magnetic moment $\boldsymbol{\mu}$ in motion acquires an electromagnetic contribution $-(\mathbf{E} \times \boldsymbol{\mu})/c^2$, where $c$ is the speed of light \cite{Moller-1955, Becker-1964, Jackson-1999}. This is a small effect which is proportional to $1/c^2$ and has been substantially overlooked in previous studies. 

In this Letter we show that, when applied to spin waves, the interaction of the magnetic moment current with the electric field corresponds to a shift of the dispersion relation in the wavenumber as $\omega(q-q_E)$ with $q_E=-\gamma_LE_x/c^2$, where $\gamma_L$ is the gyromagnetic ratio. This effect is independent from the possibly present magneto-electric effects, which modify the frequency, and superimposes to them. Therefore it is important to take it into account when the material dependent magneto-electric effects are deduced from the measured data \cite{Zhang-2014}. To conclude our study we extend our picture to arbitrary magnetization dynamics and we find that the presence of the electric field can be written in terms of a dynamic interaction. This dynamic interaction has the same chiral from of the Dzyaloshinskii-Moriya interaction but is fully tunable with the applied electric field \cite{Moon-2013}. We envisage how this effect can be possibly employed in the dynamic generation of chiral structures.

In ferromagnets, described by the continuous magnetization field $\mathbf{M}$ with constant amplitude $M_s$, the spin waves are described by the Larmor precession equation

\beq
\frac{\partial \mathbf{M}}{\partial t} = - \mu_0 \gamma_L \mathbf{M} \times \mathbf{H}_{eff}
\label{EQ:pre}
\eeq

\noindent where $\gamma_L$ is the gyromagnetic ratio \cite{Gurevich-1996, Stancil-2009}. The effective field $\mathbf{H}_{eff}$ is given by the functional derivative 
 
\beq
\mathbf{H}_{eff} = -\frac{1}{\mu_0}\frac{\delta \mathcal{U}_M}{\delta \mathbf{M}}
\label{EQ:Heffdef}
\eeq

\noindent where $\mathcal{U}_M$ is the so-called micromagnetic energy density. The energy density contains four main terms: exchange, anisotropy, magnetostatic, and applied field and is expressed as

\beq
\mathcal{U}_M = A(\boldsymbol{\nabla}\mathbf{{m}})^2+f_{AN}(\mathbf{{m}},\mathbf{{n}})- \frac{1}{2}\mu_0\mathbf{H}_M\cdot\mathbf{M}-\mu_0\mathbf{H}_a\cdot\mathbf{M}
\label{EQ:UM}
\eeq

\noindent In Eq.(\ref{EQ:UM}) the exchange term is proportional to $A$, the exchange stiffness, and $\left ( \boldsymbol{\nabla} \mathbf{{m}} \right )^{2}$ is a short-hand notation for $\left | \boldsymbol{\nabla} {m}_{x} \right |^{2} + \left | \boldsymbol{\nabla} {m}_{y} \right |^{2} + \left | \boldsymbol{\nabla} {m}_{z} \right |^{2}$ where $\mathbf{{m}} = \mathbf{M}/M_s$ is the versor of the magnetization vector. The anisotropy term is $f_{AN}$ and depends on some local easy direction $\mathbf{{n}}$. The magnetostatic field $\mathbf{H}_M$ is given by the solution of the magnetostatic equations (${\boldsymbol{\nabla} \cdot \mathbf{H}_M = - \boldsymbol{\nabla} \cdot \mathbf{M}}$ and ${\boldsymbol{\nabla} \times \mathbf{H}_M=0}$) and $\mathbf{H}_a$ is the applied field. By performing the functional derivative of Eq.(\ref{EQ:Heffdef}) one obtains the classical expression for the effective field

\beq
\mathbf{H}_{eff}= l_{EX}^{2} \boldsymbol{\nabla}^{2} \mathbf{M} - \frac{\partial f_{AN}}{\partial\mathbf{M}} + \mathbf{H}_{M} + \mathbf{H}_{a}
\label{EQ:Heff}
\eeq

\noindent where $l_{EX} = [2A/(\mu_0M_s^2)]^{1/2}$ is the exchange length \cite{Gurevich-1996, Stancil-2009}. The spin waves are the solutions of Eq.(\ref{EQ:pre}) with Eq.(\ref{EQ:Heff}) when the magnetization field is decomposed as the sum of a large static component $\mathbf{M}_{\|}$ plus a small time dependent one $\mathbf{M}_{\bot}(t)$ as $\mathbf{M} = \mathbf{M}_{\|}+\mathbf{M}_{\bot}(t)$ and Eq.(\ref{EQ:pre}) is linearized. 

As spin waves can have a group velocity they can contribute to the transport of extensive quantities. Specifically they can transport magnetic moment \cite{Maekawa-2017}. However, because of an electromagnetic effect, the transport of a magnetic moment corresponds to an electric polarization. Then a static electric field $\mathbf{E}$ will have an effect. Indeed, a point particle with magnetic moment $\boldsymbol{\mu}$ in motion with velocity $\mathbf{v}$ has, in the laboratory frame, an electric dipole moment $\mathbf{p} = \gamma_v \mathbf{\mu}\times\mathbf{v}/c^2$, where $\gamma_v = \left[1-(v/c)^2\right]^{-1/2}$ is the Lorentz factor. This property is the direct consequence of the Lorentz transformations for electromagnetic fields and sources between different inertial reference frames \cite{Jackson-1999, Becker-1964, Moller-1955}. In presence of a static electric field the corresponding energy term is $\mathcal{U}_P = -\mathbf{E} \cdot \mathbf{P}$ that, at low velocity, $v \ll c $ (i.e. $\gamma_v\simeq 1$), reads $- \mu_0 \boldsymbol{\mu}\cdot \mathbf{v}\times\mathbf{E}/c^2$. However, this term, which explicitly contains the velocity of the particle, does not change the energy of the system but changes its momentum. This is readily seen by taking the Lagrangian of the magnetic moment in motion, $\mathcal{L} = \mathcal{T}-\mathcal{U}_P-\mathcal{U}_M$, and deriving the canonical momentum, $\mathbf{\pi} = \partial \mathcal{L}/\partial \mathbf{v}$, which turns out to be $\boldsymbol{\pi} = m^*\mathbf{v} - (\mathbf{E} \times \boldsymbol{\mu})/c^2$ i.e. the sum of the kinetic contribution ($m^*$ is the mass of the particle) and the electromagnetic contribution which is due to the presence of the electric field \cite{Fisher-1971}. By writing the Hamiltonian

\beq
\mathcal{H} = \frac{1}{2m^*} \left( \boldsymbol{\pi} + \frac{1}{c^2} (\mathbf{E} \times \boldsymbol{\mu}) \right)^2 - \mu_0 \boldsymbol{\mu}\cdot \mathbf{H}
\label{EQ:Ham_par}
\eeq

\noindent it becomes clear that the total energy has not changed. The presence of the electromagnetic momentum is also at the base of the quantum mechanical interference of neutral particles with magnetic moment, the so-called Aharonov-Casher effect \cite{Anandan-1982, Aharonov-1984}. Indeed, in wave mechanics, the electromagnetic momentum leads to the accumulation of a phase $\varphi =  - [\int_{\mathcal{C}} ( \mathbf{E} \times \boldsymbol{\mu} ) \cdot d\mathbf{l}]/({\hslash c^2})$ over the path $\mathcal{C}$ that can give rise to quantum interference for closed paths \cite{Aharonov-1984, AlJaber-1991}. 
 
Returning back to our spin wave problem, we must be aware that whenever the spin waves generate a current density of magnetic moment, $\mathbf{j}_{\mathbf{M}}$, we will have an electric polarization $\mathbf{P}$ in the laboratory frame. If we express the tensor of the magnetic moment current as the product $\mathbf{j}_{\mathbf{M}} =  \mathbf{v}_g \mathbf{M}_{sw}$ of the group velocity $\mathbf{v}_g$ and the magnetization carried by the spin wave $\mathbf{M}_{sw}$, the electric polarization will be $\mathbf{P} = \mathbf{v}_g \times \mathbf{M}_{sw} /c^2$ and the additional energy term will be $- \mathbf{E}\cdot ( \mathbf{v}_g \times \mathbf{M}_{sw}) /c^2$. Such a term is not present in the micromagnetics' equations ((\ref{EQ:pre}) and (\ref{EQ:Heff})) and it is also not so obvious how to introduce it  \cite{Cao-1997} because both $\mathbf{v}_g$ and $\mathbf{M}_{sw}$ will be given by the solution of the spin wave problem.

Here we propose a method to answer to this question by using the Lagrangian approach to micromagnetism \cite{Brown-1963}. In absence of electric field the Lagrangian of micromagnetism $\mathcal{L}_0$ is given by

\beq
\mathcal{L}_0 = \mathcal{T} - \mathcal{U}_M
\eeq

\noindent where $\mathcal{U}_M$ is the micromagnetic energy density of Eq.(\ref{EQ:UM}) and $\mathcal{T}$ is the kinetic term. It can be shown that with $\mathcal{T} = (M_{s}/\gamma_L) \mathbf{n}\times\mathbf{m}\cdot\dot{\mathbf{m}}/(1+\mathbf{n}\cdot\mathbf{m})$, where $\mathbf{n}$ is an arbitrary unit vector, the Euler-Lagrange equations will reproduce the precessional equation of motion of Eq.(\ref{EQ:pre}). With $\mathbf{n} = \hat{\mathbf{z}}$ we use the spherical coordinates for the magnetization unit vector (${m}_x=\sin\theta\cos\varphi$, ${m}_y=\sin\theta\sin\varphi$ and ${m}_z=\cos\theta$) and obtain

\beq
\mathcal{T} = \frac{M_s}{\gamma_L} (1-\cos\theta)\dot{\varphi}
\label{EQ:LT}
\eeq

\noindent In presence of the static electric field, the term $-\mathcal{U}_P$ must be added to the Lagrangian $\mathcal{L}_0$ so that the full Lagrangian is $\mathcal{L} = \mathcal{L}_0-\mathcal{U}_P$. However to explicit $\mathcal{U}_P$  we need an expression for $\mathbf{j}_{\mathbf{M}}$. By taking the main magnetization along $z$, the velocity along $y$ and static electric field along $x$ (see Fig.1) we have

\beq
j_{M,yz} = v_{g,y} M_{sw,z}
\eeq

\noindent For small oscillations of the magnetization vector around the $z$ axis we take the density of magnetic moment transported by the spin wave as

\beq
M_{sw,z} = - M_s \frac{1}{2} \theta^2
\eeq

\noindent and by expressing the $\varphi$ angle for a plane wave as ${\varphi = \omega t - q_y y+\varphi_0}$, where $\omega = \dot{\varphi}$ and $q_y = -\nabla_y \varphi$, we obtain the group velocity as

\beq
v_{g,y} = \frac{\partial \omega}{\partial q_y} = - \frac{\partial \dot{\varphi}}{\partial \nabla_y \varphi} 
\eeq

\noindent The group velocity is well defined when both $\omega$ and $q_y$ are slowly varying parameters, i.e. their time and space dependent is much slower than those of $\varphi(y,t)$ \cite{Whithams-1999}. For small oscillations of the magnetization ($\theta \ll 1$) the kinetic term of the Lagrangian is approximated as $\mathcal{T} \simeq ({M_s}/{\gamma_L}) (\theta^2/2)\dot{\varphi}$, so we obtain the full Lagrangian

\beq
\mathcal{L} = \frac{1}{2} \theta^2 \frac{M_s}{\gamma_L} \left[ \dot{\varphi} + \frac{\partial \dot{\varphi}}{\partial \nabla_y \varphi}  \frac{\gamma_L E_x}{c^2}  \right]  - \mathcal{U}_M
\label{EQ:L}
\eeq

\noindent We now observe that the derivative ${\partial \dot{\varphi}}/{\partial \nabla \varphi}$ will be known only once one has the solution, i.e. the dispersion relation $\omega(q_y)$. However, even without having the explicit form, we note that the dependence $\omega(q_y)$ will denote a particular property of the solution and that the term under squared parenthesis of Eq.(\ref{EQ:L}) can be taken as the first order Taylor expansion of ${\omega(q_y - {\gamma_L E_x}/{c^2})}$ around $E_x=0$, i.e.

\beq
\omega\left(q_y - \frac{\gamma_L E_x}{c^2}  \right) \simeq \omega(q_y) - \frac{\partial \omega}{\partial q_y} \frac{\gamma_L E_x}{c^2}
\eeq

\noindent The role played by the electric field is therefore those of changing the wavenumber as

\beq
q_y \rightarrow q_y +  \frac{\gamma_L E_x}{c^2}
\label{EQ:qprime}
\eeq

\noindent In terms of the phase $\varphi$ this corresponds to the substitution of the derivative operator $\nabla_y\varphi \rightarrow D_y\varphi $ with

\beq
D_y\varphi = \nabla_y \varphi -  \frac{\gamma_L E_x}{c^2}
\label{EQ:nabla_red}
\eeq

\noindent By operating this change the full Lagrangian of Eq.(\ref{EQ:L}) becomes formally identical to the original Lagrangian $\mathcal{L}_0$, i.e. those for the precession without the electric field. As from the original Lagrangian $\mathcal{L}_0$ we can solve the spin wave problem and find the dispersion relations with $E_x=0$, this means that the dispersion relations with the static electric field will be given by taking the known equations and applying the redefinition of the derivative $\nabla_y\varphi \rightarrow D_y\varphi$ of Eq.(\ref{EQ:nabla_red}).

To make a first example we consider exchange spin waves in which we disregard the magnetostatic field. The expression for the dispersion relation is $\omega = \omega_H + \omega_M l_{EX}^2 (q_y)^2$ where $\omega_H = \mu_0 \gamma_L H_z$ and $\omega_M = \mu_0 \gamma_L M_s$ \cite{Stancil-2009}. Using Eq.(\ref{EQ:qprime}) we obtain that the dispersion relation in the static electric field as

\beq
\omega = \omega_H + \omega_M l_{EX}^2 \left(q_y +\frac{\gamma_L E_x}{c^2} \right)^2
\label{EQ:omegaEX}
\eeq

\noindent which is shifted along the axis of the wavenumber of the quantity $-{\gamma_L E_x}/{c^2}$. To verify that this expression describes the physics of the problem we make the quantization of the spin waves of Eq.(\ref{EQ:omegaEX}) in order to compare it with the Hamiltonian of the particle of Eq.(\ref{EQ:Ham_par}). The energy of the quantum of the spin wave (magnon) is $\epsilon = \hslash \omega$ and its elementary magnetic moment is $\mu_z = -2\mu_B$ ($\mu_B$ is the Bohr magneton). We then obtain

\beq
\epsilon = \frac{1}{2m^*} \left( \hslash q_y + \frac{1}{c^2} E_x \mu_z \right)^2 - \mu_0 \mu_z H_z 
\label{EQ:Energy_exchange_magnon}
\eeq

\noindent where $m^* = \hslash^2M_s/(8A\mu_B)$ is the effective mass of exchange spin waves. By taking the vector magnetic moment $\boldsymbol{\mu} = \mu_z \hat{\mathbf{z}}$, the vector momentum $\boldsymbol{\pi} = \hslash q_y \hat{\mathbf{y}}$ and the vector field $\mathbf{E} = E_x \hat{\mathbf{x}}$ we find that the energy of the magnon in an electric field is identical to the Hamiltonian of Eq.(\ref{EQ:Ham_par}) \cite{Meier-2003}.

\begin{figure}[htb]
\centering
\includegraphics[width=8cm]{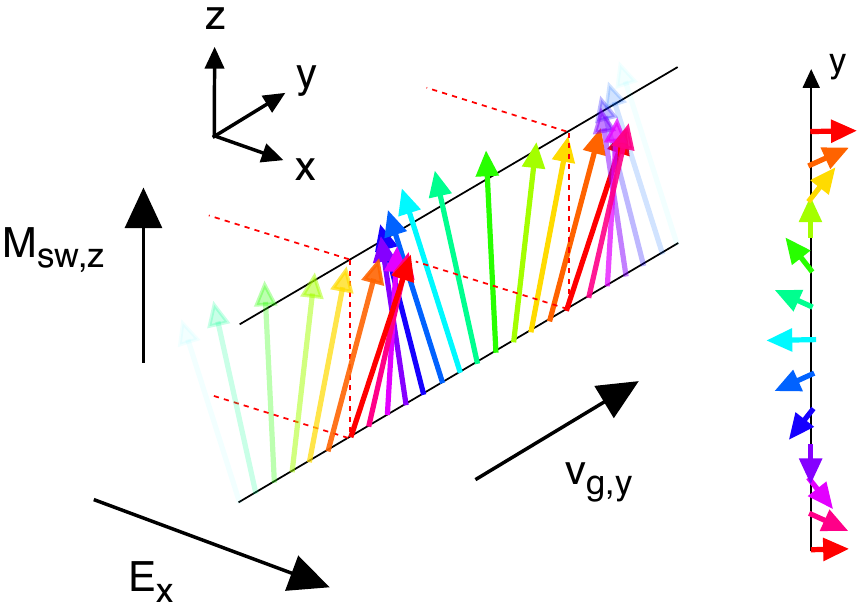}
\caption{Spin waves in an electric field. The ferromagnet has main magnetization along $z$. The spin waves are the small oscillations of the magnetization vector around the $z$ axis (top view at the right hand side). The magnetic moment current transported by the spin wave is $j_{M,yz} = v_{g,y} M_{sw,z}$ where $v_{g,y}$ is the group velocity (chosen along $y$). The static electric field $E_x$ is along $x$.} \label{FIG:1}
\end{figure}

As a different example we test the possibility to tune the propagation of magnetostatic waves. By disregarding the exchange and taking a finite shape for the magnetic body one finds that the linearized spin wave problem can be solved by using the Walker's equation ${\boldsymbol{\nabla}^2 \Phi = - \boldsymbol{\nabla} \cdot \bar{\chi} \boldsymbol{\nabla} \Phi}$ for the time harmonic part of the magnetostatic potential $\Phi$ ($\bar{\chi}$ is the susceptibility tensor) \cite{Gurevich-1996, Stancil-2009}. In presence of a static electric field we have to apply the redefinition of the derivative operator of Eq.(\ref{EQ:nabla_red}). In the case of the Walker's equation, the derivative operates over complex vectors (i.e. the magnetization vector is $\mathbf{M}_{\bot}=\mathbf{M}_{\bot,0}\exp(iq_y y)$ where $\mathbf{M}_{\bot,0}$ is a complex amplitude) therefore Eq.(\ref{EQ:nabla_red}) is generalized as

\beq
\mathbf{D} =  \hat{\mathbf{x}} \nabla_x + \hat{\mathbf{y}} \left(  \nabla_y + i \frac{\gamma_LE_x}{c^2}  \right)  +  \hat{\mathbf{z}}  \nabla_z
\label{EQ:nablam}
\eeq

\noindent and the Walker's equation becomes $\mathbf{D}^2 \Phi = - \mathbf{D} \cdot \bar{\chi} \mathbf{D} \Phi$. For a thin film of thickness $d$ along $x$ and magnetized along $z$ we have surface waves traveling along $y$. The dispersion relation is then

\beq
\omega = \sqrt{ \omega_0^2 + \frac{\omega_M^2}{4} \left[ 1- \exp\left(-2\left|q_y +  \frac{\gamma_L E_x}{c^2}\right|d\right) \right]}
\label{EQ:surf}
\eeq

\noindent where $\omega_0 = \sqrt{\omega_H (\omega_H + \omega_M)}$. Again, the effect of the static electric field $E_x$ is a linear shift of the dispersion relation. Then for a spin wave with $q_y>0$ the presence of a positive electric field $E_x>0$ will correspond to an increase of the frequency. At $\omega \simeq \omega_0$  the dispersion relation is approximated as 

\beq
\omega \simeq \omega_0 + v_{g} \left|q_y +  \frac{\gamma_L E_x}{c^2}\right|
\eeq

\noindent where $ v_{g} = {\omega_M^2d}/{(4 \omega_0)}$  is the group velocity. Then the relative frequency change due to the electric field $E_x$ with respect to $E_x=0$ is

\beq
\frac{\Delta \omega}{\omega_0} = \frac{\omega_M^2}{4 \omega_0^2}\frac{\gamma_L}{c^2} d \, E_x
\eeq

\noindent With the gyromagnetic ratio for the electron spin $\gamma_L \simeq 1.761 \cdot 10^{11}$ s$^{-1}$T$^{-1}$ and using the speed of light in vacuum, we find the coefficient $\gamma_L/c^2 = 1.95 \cdot10^{-6}$ V$^{-1}$. The effect can be observable and possibly exploitable with $H_z \ll M_s$. With yttrium iron garnet (YIG) as magnetic material ($\mu_0M_s \simeq 0.18$ T) and by taking $\mu_0H_z \simeq 1\cdot10^{-3}$ T we have $f_0 \simeq 0.38$ GHz and a relative shift of $8.8\cdot10^{-5}$ V$^{-1}$. This means a frequency change of the order 0.1\% for $d \, E_x =$12 V. These numbers should be readably observable in specific resonance experiments with thin magnetic films. In order to compare the effect of the joint presence of the material dependent magneto-electric effect of Refs.\cite{Mills-2008, Liu-2011, Wang-2018, Krivoruchko-2018} and our magnetic moment current one, we better compute the electric field induced phase per unit length $\Delta\varphi/\Delta y$. By taking the magneto-electric energy term as ${\mu_0 (b/2) \mathbf{E} \cdot \left[ \mathbf{M}(\boldsymbol{\nabla}\cdot\mathbf{M}) - (\mathbf{M}\cdot\boldsymbol{\nabla}) \mathbf{M}\right]}$ \cite{Wang-2018}, the result is ${\Delta\varphi/\Delta y = [ \gamma_L/c^2 + (\omega-\omega_0) \omega_M b /v_g^2] E_x}$ which shows that the induced phase has two component: the magnetic moment current one is independent of the frequency while the magneto-electric one is frequency dependent. Two terms of this kind are actually seen in the experimental data \cite{Zhang-2014} and the predicted  frequency independent phase ($\gamma_L/c^2 E_x$) is of the same order of magnitude of the measured one. Our result points therefore toward to critical reconsideration of the interpretation of the existing literature data in terms of these two concurrent effects \cite{Ansalone-2020}.

\begin{figure}[htb]
\includegraphics[width=8cm]{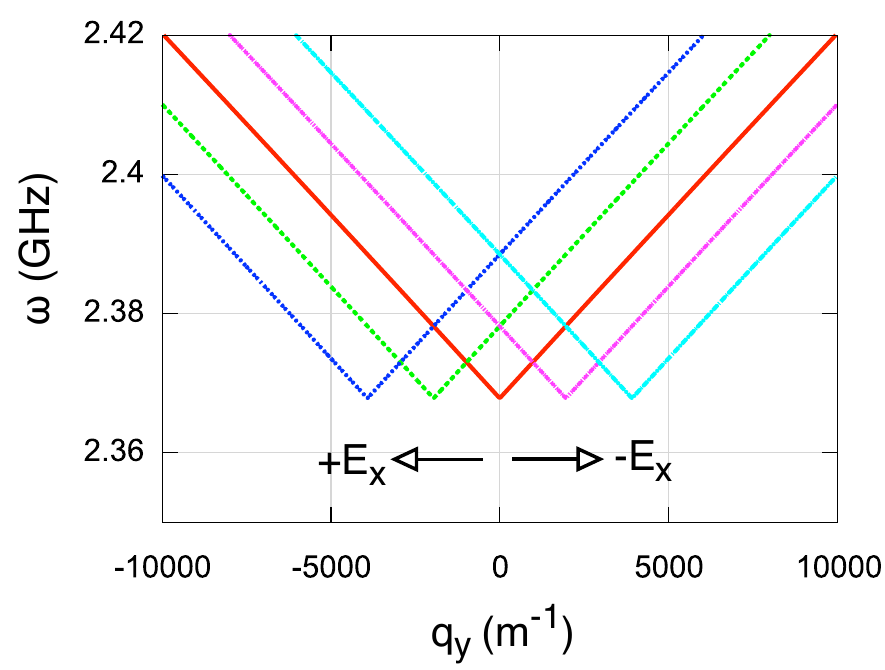}
\caption{Effect of the electric field $E_x$ on the spin wave dispersion relation $\omega(q_y)$. The plot is Eq.(\ref{EQ:surf}) for magnetostatic surface waves with film thickness $d=50$ nm, $\mu_0M_s = 0.18$ T (YIG), $\mu_0H_z \simeq 1\cdot10^{-3}$ T. In the plot the curves are at electric potential $d \, E_x = 0; \pm 50$ V and $\pm 100$ V.} \label{FIG:2}
\end{figure}

Having derived the change in the behavior of spin waves because of an applied electric field we wonder now how to extend the picture to arbitrary magnetization dynamics. The Lagrangian approach is attractive from this point of view because it is also the appropriate framework to incorporate damping effects and have a full theory for the dynamics \cite{Bertotti-2009}. To do this extension we make the following two steps. 1) We assume that, for large magnetization deviations, the magnetic moment current is $j_{M, yz} = (1-\cos\theta) M_s {\partial \dot{\varphi}}/{\partial \nabla_y \varphi}$, i.e. the amplitude of the transported magnetization has the same functional form of the kinetic term of the Lagrangian (Eq.(\ref{EQ:LT})). This is a fully reasonable assumption once we expect that the transported moment is proportional to $-(1-{m}_z)$. 2) We extend the vector operator $\mathbf{D}$ of Eq.(\ref{EQ:nablam}) to vector magnetization and vector electric field. The transformation of each component of the differential operator over each component of the magnetization unit vector is $\nabla_i { m}_j \rightarrow D_i { m}_j $ with

\beq
D_i { m}_j =  \nabla_i { m}_j + \frac{\gamma_L}{c^2} \left[ E_jm_i - \mathbf{E}\cdot\mathbf{m} \,\,\, \delta_{ij}\right]
\label{EQ:nabla_vec}
\eeq

\noindent Now, once again, if we operate the transformation ${\boldsymbol{\nabla} \rightarrow \mathbf{D}}$, we find that the full Lagrangian is formally transformed back to $\mathcal{L}_0$. But now the micromagnetic energy density $\mathcal{U}_M$ is expressed as a function of $\mathbf{D} \mathbf{m}$. Even without writing the full dynamic equation, it is of interest to look at the form taken by the micromagnetic energy terms after this transformation. Only the terms containing the derivatives are affected, i.e. exchange and magnetostatic. We take the exchange energy as an example. With $\mathcal{U}^{\prime}_{EX} = A ( \mathbf{D} \mathbf{ m})^2 $ we find

\begin{multline}
\mathcal{U}^{\prime}_{EX} = A( \boldsymbol{\nabla} \mathbf{m})^2 \\ - 2A \frac{\gamma_L}{c^2} \mathbf{E}\cdot \left[ \mathbf{m}(\boldsymbol{\nabla}\cdot\mathbf{m}) - (\mathbf{m}\cdot\boldsymbol{\nabla}) \mathbf{m}\right] \\ +A \left( \frac{\gamma_L}{c^2} \right)^2\left[ E^2 + (\mathbf{E}\cdot \mathbf{m})^2\right]
\label{EQ:Uexp}
\end{multline}

\noindent At the right hand side we recognize the usual exchange, $\mathcal{U}_{EX} = A( \boldsymbol{\nabla}\mathbf{ m})^2$, plus two additional terms that describe the dynamic coupling with the electric field. They are present because the system, under its dynamic evolution, is generating a transport of magnetic moment in space. The fact that the time derivative of the magnetization unit vector, $\dot{\mathbf{ m}}$, is not explicit in the two new terms of the the expression (\ref{EQ:Uexp}), does not mean that these are static terms. The two terms are not present in the static theory describing the energy of the micromagnetic stable states (in which the energy is still given by Eq.(\ref{EQ:UM})), but appear as the result of the dynamic interaction between the transport of magnetic moment and the static electric field. The second term at the right hand side of Eq.(\ref{EQ:Uexp}) is the Lifshitz invariant corresponding to the Dzyaloshinskii-Moriya (DM) interaction in continuous form and to the magneto-electric coupling \cite{Mostovoy-2006, Dzyaloshinskii-2008, Moon-2013}. The third term is an anisotropy energy (an easy plane ($x,y$) type if the electric field is along $z$). The second term gives a dynamic interaction which is chiral and brings therefore an unexpected light over the possibility to develop chiral structures in ferromagnets by applying an external electric field even in absence of intrinsic chiral effects of magneto-eletric or spin-orbit origin. The electric field can be effective in the formation of specific static chiral configurations, because it may contribute to choose a specific minimum rather another one when the system relaxes from an high energy state toward an energy minimum. The sign of the applied electric field could be then a flexible method to select one type of chiral structure rather than another one by selecting a specific path of dynamic evolution. Even if the chiral dynamic interaction may be possibly smaller than the DM interaction in thin films \cite{Moon-2013} it will be interesting to derive the explicit magnetization paths followed by the system relaxation in presence of electric field. 

In conclusion we have studied the interaction between the magnetic moment current transported by the spin waves and a static electric field. The interaction we are interested in is the relativistic effect for which a magnetic moment in motion corresponds to an electric dipole. This is an effect of the order $1/c^2$, which has been largely overlooked up to now because it was believed to be too small. However we have shown that the induced phase  is of the same order of magnitude of the those induced by magneto-electric coupling in centrosymmetric ferrites \cite{Liu-2011, Zhang-2014}. By working within a Lagrangian approach we have shown that the Lagrangian of the ferromagnet in the electric field is transformed into the classical Lagrangian upon the redefinition of the differential operator. This directly shows that the shift in the wavenumber works for any kind of interaction giving rise to a group velocity for the spin waves, i.e. not only for exchange but also for magnetostatics. By further extending the picture to arbitrary magnetization dynamics we found that the electric field appears as a dynamic interaction terms  which are essentially how the energy of the system is seen when the system state is still very far from the energy minima. The main of these terms has exactly the same functional form of the Dzyaloshinskii-Moriya interaction, therefore it is chiral. Most probably the dynamic interaction can result to be much smaller then the usual Dzyaloshinskii-Moriya interaction in thin films or bulk materials \cite{Moon-2013}. However what is interesting here is that the electric field is breaking the symmetry of the problem in exactly the same way.


\end{document}